\documentclass[preprint,12pt]{elsarticle}
\usepackage{amsmath}
\usepackage{amssymb}
\usepackage{amsfonts}
\usepackage{amscd}
\usepackage[all]{xy}

\journal{Journal Physics Letters A}

\begin{document}

\begin{frontmatter}

\title{New 1-step extension of the Swanson oscillator and superintegrability of its two-dimensional generalization}

\author{Bijan Bagchi$^{1,}$\thanks{Electronic address: bbagchi123@gmail.com} \ and Ian Marquette$^{2,}$\thanks{Electronic address: i.marquette@uq.edu.au}\\
{\small\sl $^1$ Department of Appplied Mathematics, University of Calcutta,}\\
{\small \sl 92 Acharya Prafulla Chandra Road, Kolkata 700 009, India}\\
{\small\sl $^2$ School of Mathematics and Physics, The University of Queensland,} \\
{\small \sl Brisbane, QLD 4072, Australia}}
\date{ }

\begin{abstract}
We derive a one-step extension of the well known Swanson oscillator that describes a specific type of pseudo-Hermitian quadratic Hamiltonian connected to an extended harmonic oscillator model. Our analysis is based on the use of the techniques of supersymmetric quantum mechanics and address various representations of the ladder operators starting from a seed solution of the harmonic oscillator expressed in terms of a pseudo-Hermite polynomial. The role of the resulting chain of Hamiltonians related via similarity transformation is then exploited. In the second part we write down a two dimensional generalization of the Swanson Hamiltonian and establish superintegrability of such a system.
\end{abstract}




\end{frontmatter}

\newpage
%
%
\section{Introduction}

Seeking rational extensions of known solvable systems is currently one of the major topics of research in quantum mechanics. Indeed, recent times have witnessed a great deal of activity in generating rationally extended models in the context of polynomial Heisenberg algebras (PHA) and exceptional orthogonal polynomials \cite{Que1,Mar0,Fer1,Car1}. A particularly simple one is  for the problem of harmonic oscillator in whose case the ladder operators have been constructed through the combination of the oscillator creation and annihilation operators along with the supercharges or having the combination of the latter. In this regard, harmonic rational extension have been carried out in the framework of a supersymmetric quantum mechanics (SUSYQM) theory \cite{MarQue, Mar1}. In particular, 1-step extensions have been sought for the radial oscillator and its generalization, the Scarf I ( also sometimes referred to as Poschl-Teller or Poschl-Teller I) \cite{Que2,Que3,Oda1,Oda2,Oda3,Sas1,Gra1,Ho1,Yad}, and the generalized Poschl-Teller ( also sometimes referred to as hyperbolic Poschl-Teller or Poschl-Teller II) \cite{Bag1,Oda1,Oda2}. \par
%
%
In this article we apply standard supersymmetric (SUSY) techniques to develop a systematic procedure for obtaining a solvable rational extension of a non-Hermitian quadratic Hamiltonian that was proposed by Swanson \cite{Swa1} sometime ago and was shown to possess real and positive eigenvalues. Such a model that we present in its differential operator form is new and open the way of constructing new classes of non-Hermitian quadratic Hamiltonians based on rational extensions. We also address a general two-dimensional analog of Swanson Hamiltonian from a two-dimensional perspective that is separable in Cartesian coordinates and establish superintegrability of such a system by making suitable use a underlying ladder operators.

\section{Swanson model}

Swanson model deals with a specific type of a non-Hermitian Hamiltonian connected to an extended harmonic oscillator problem. A general quadratic form for it has the simple structure
\begin{equation}
H^{s}=\omega a^{\dagger}a +\alpha a^{2} +\beta (a^{\dagger})^{2} +\frac{1}{2}\omega \label{HS}
\end{equation}
where $a$ and $a^{\dagger}$ are respectively the usual annihilation and creation operators of the one-dimensional harmonic oscillator obeying the canonical commutation relation $[a,a^{\dagger}]=1$. In (\ref{HS}), $\omega$, $\alpha$ and $\beta$ are real constants. It is clear that with $\alpha \neq \beta$, $H^{s}$ ceases to be Hermitian. Nonetheless, it is pseudo-Hermitian \cite{Mos2} embodying parity-time symmetry and supports a purely real, positive spectrum over a certain range of parameters. Swanson Hamiltonian has been widely employed as a toy model to investigate a wide class of non-Hermitian systems for different situations. Some of its applications have been in exploring the choice of a unique and physical metric operator to generate an equivalent Hermitian system \cite{Jones,Mus,Bag2,Mos1,Mos2}, setting up of a group structure of the Hamiltonian \cite{Qes1,Asi}, looking for possible quasi-Hermitian and pseudo-supersymmetric (SUSY) extensions \cite{Ban,Qes2,Yes1}, working out a $\mathcal{N}$-fold SUSY connection \cite{BT}, estimating minimum length uncertainty relations that result from the structure of non-commutative algebras \cite{Fri1,Fri2}, studying a relevant R-deformed algebra \cite{Rr}, writing down supercoherent states \cite{CDMT} and investigating some of the aspects of classical and quantum dynamics for it \cite{DG,Gra}.

 \par
%
%
A similarity transformation allows us to write down \cite{Jones} the Hermitian equivalence of $H^{s}$. In this way an equivalent Hermitian counterpart
of $H^{s}$ turns out to be a scaled harmonic oscillator :
\begin{equation*}
h=\rho H \rho^{-1}
\end{equation*}
\begin{equation}
= -\frac{1}{2}  (\omega-\alpha -\beta )\frac{d^{2}}{dx^{2}} +\frac{1}{2}x^{2} \frac{\omega^{2}-4\alpha \beta}{\omega -\alpha -\beta} +\frac{1}{2}\omega  \label{h}
\end{equation}
where $\rho=e^{\frac{1}{2}\lambda x^{2}}$ along with the accompanying eigenfunctions
\begin{equation}
\psi_{n}(x)= N_{n} e^{-\frac{1}{2}x^{2}(\lambda +\Delta^{2})} H_{n}(\Delta x). \label{eig}
\end{equation}
In (\ref{h}) and (\ref{eig}) the parameters $\lambda$ and $\Delta$ are defined by
\begin{equation}
\lambda=\frac{\beta -\alpha}{\omega-\alpha-\beta},\quad \Delta = \frac{(\omega^{2}-4 \alpha \beta)^{\frac{1}{4}}}{(\omega -\alpha-\beta)^{\frac{1}{2}}}
\end{equation}
and $H_{n}$ is a nth degree Hermite polynomial. The eigenfunctions are orthonormal with respect to the quantity $e^{\lambda x^{2}}$ i.e.
\begin{equation}
\int \psi_{m}^{*}(x)e^{\lambda x^{2}} \psi_{n}(x)dx=\delta_{mn}.
\end{equation}
\par
%
%
We note that the scaled harmonic oscillator Hamiltonian $h$ as given by (\ref{h})
can be cast into the standard form through the transformation
\begin{equation}
h \rightarrow \tilde{h}=\frac{2}{\sqrt{\omega^{2}-4 \alpha \beta }} ( h -\frac{1}{2}\omega)
\end{equation}
and introducing a change of variable $x \rightarrow z$ as given by
\begin{equation}
x \rightarrow z =\sqrt[4]{\frac{\omega^{2}-4 \alpha \beta}{(\omega - \alpha -\beta)^{2}}} x \equiv \Delta x .
\end{equation}
Thus we arrive at the following Schr\"{o}dinger operator for $\tilde{h}$ :
\begin{equation}
\tilde{h}=-\frac{d^{2}}{dz^{2}}+z^{2}  \label{ht}
\end{equation}
from which we develop a SUSY scheme by means of standard supercharges that go with it.

\section{SUSY scenario}

The Hamiltonian $\tilde{h}$ can be embedded in a supersymmetric setting \cite{Mar1}
by defining a pair of partner Hamiltonians in terms of the z-coordinate
\begin{equation}
\tilde{h}^{(+)}=\tilde{A}^{\dagger}\tilde{A}=-\frac{d^{2}}{dz^{2}}+\tilde{V}^{(+)}-\tilde{E} \equiv \tilde{h} -\tilde{E} \label{htp}
\end{equation}
\begin{equation}
\tilde{h}^{(-)}=\tilde{A}\tilde{A}^{\dagger}=-\frac{d^{2}}{dz^{2}}+\tilde{V}^{(-)}-\tilde{E}
\end{equation}
where the operators $\tilde{A}$ and $\tilde{A}^{\dagger}$ are governed by the superpotential $\tilde{W}(z)$:
\begin{equation}
\tilde{A}^{\dagger}=-\frac{d}{dz}+\tilde{W}(z),\quad \tilde{A}=\frac{d}{dz}+\tilde{W}(z).
\end{equation}
This provides identification of the corresponding partner potentials
\begin{equation}
\tilde{V}^{(\pm)}(z)=\tilde{W}^{2}(z)\mp \tilde{W'}(z) + \tilde{E}.
\end{equation}
\par
%
%
It should be remarked that the component Hamiltonians $\tilde{h}^{(+)}$ and $\tilde{h}^{(-)}$ intertwine with the operators $\tilde{A}$ and $\tilde{A}^{\dagger}$ in the manner $\tilde{A}\tilde{h}^{(+)}=\tilde{h}^{(-)}\tilde{A}$ and $\tilde{A}^{\dagger}\tilde{h}^{(-)}=\tilde{h}^{(+)}\tilde{A}^{\dagger}$. Further the underlying nodeless eigenfunction $\tilde{\phi}(x)$ of the Schr\"{o}dinger equation
\begin{equation}
(-\frac{d^{2}}{dz^{2}}+\tilde{V}^{(+)})\tilde{\phi}(z)=\tilde{E}\tilde{\phi}(z) \label{zhamil}
\end{equation}
has the feature that it is given by $\tilde{W}(z)=-(log(\tilde{\phi}(z))'$ where the prime stands for the derivative with respect
to z. The factorization energy $\tilde{E}$ is assumed to be smaller than or equal to the ground-state energy of $\tilde{V}^{(+)}$. From
(\ref{ht}) and (\ref{htp}), it is clear that $\tilde{V}^{(+)}$ is identifiable with $\tilde{V}^{(+)}=z^{2}$. Then with $\tilde{E}=1$ and $\tilde{W}(z)=z$, the partner potential turns out to be $\tilde{V}^{(-)}=z^{2}+2$ reflecting the shape-invariance character of the harmonic oscillator, $\tilde{V}^{(-)}$ being just a translated oscillator with respect to $\tilde{V}^{(+)}$.
\par
%
%
If however, $\tilde{E}<1$, then the only possible nodeless seed solutions of (\ref{zhamil}) are of the type
\begin{equation}
\tilde{\phi}_{m}(z)=\mathcal{H}_{m}(z)e^{\frac{1}{2}z^{2}}, \quad m=2,4,6,...  \label{seed}
\end{equation}
where the pseudo-Hermite polynomial $\mathcal{H}_{m}(z)$ is related to the standard Hermite by
$\mathcal{H}_{m}(x)=(-1)^{m}H_{m}(ix)$.
We remark that $\psi_{m}$ of the equivalent Hamiltonian representation of Swanson Hamiltonian is invariant under
$x \rightarrow ix$. This means that given the correspondence between the Hermite polynomials and their pseudo-Hermite counterparts, it follows from (\ref{eig}) that
\begin{equation}
\psi_{n}(ix)=N_{n}e^{\frac{1}{2}x^{2}(\lambda +\Delta^{2})}\mathcal{H}_{n}(i\Delta x),\quad n=0,1,2,3,...
\end{equation}
is also an eigenstate of $H^{s}$. Now transforming $\psi_{n}(ix)$ as $\psi_{n}(ix) \rightarrow \rho^{-1}\psi_{n}(ix)$ we in fact recover the seed solution $\psi_{n}(ix)=N_{n}e^{\frac{1}{2}x^{2}(\lambda +\Delta^{2})}H_{n}(ix)$ where $H_n(ix)$ is the standard Hermite polynomial. \par
%
%
Given $\tilde{\phi}_{m}(z)$ as (\ref{seed}), the accompanying superpotential $\tilde{W}(z)$ is given by
\begin{equation}
\tilde{W}(z)=-z-\frac{\mathcal{H}'_{m}}{\mathcal{H}_{m}}\quad (  \equiv-\frac{\tilde{\phi'}}{\tilde{\phi}} ). \label{Wtz}
\end{equation}
The partner potential $\tilde{V{(-)}}(z)$ then reads
\begin{equation}
\tilde{V}^{(-)}(z) =z^{2} -2 [ \frac{\mathcal{H}_{m}''}{\mathcal{H}_{m}} -   (\frac{ \mathcal{H}_{m}'}{\mathcal{H}_{m}}  )^{2} +1 ] \label{Vm}
\end{equation}
along with the energy spectrum
\begin{equation}
\tilde{E}_{m}=-2m-1.
\end{equation}
The explicit forms first appear in \cite{Fel} in the course of deriving the exact closed form solutions of a generalized one-dimensional potential \cite{Car} that has a form intermediate to the harmonic and isotonic oscillators. Indeed the latter corresponds to $m=2$ which is the second category of rational extension. Subsequently, a translational shape invariant property of (\ref{Wtz}) with a zero translational amplitude was established in \cite{Gran} and also more recently discussed in the Krein-Adler and Darboux-Crum construction of these systems \cite{Gom,Oda4}.\par
%
%
At this stage it is instructive to revert to the x-coordinate and write down the supersymmetric partner Hamiltonian counterpart of $H^{s}$. Defining
\begin{equation}
J=\frac{2}{\sqrt{\omega^{2}-4\alpha \beta}}, z=\Delta x,\quad E=\frac{\tilde{E}}{J}
\end{equation}
and noting that $\tilde{h}^{(+)}$ represents the form of (\ref{h}), the $\tilde{h}^{(+)}$ component reads in terms
of the parameters $J$ and $\Delta$ :
\begin{equation}
h^{(+)}=-\frac{1}{J\Delta^{2}}\frac{d^{2}}{dx^{2}}+\frac{\Delta^{2}}{J}x^{2}+\frac{\omega}{2}-E.
\end{equation}
\par
%
%
It is straightforward to obtain from (\ref{Vm}) the form for $h^{(-)}$ namely
\begin{equation}
h^{(-)}=\frac{1}{J \Delta^{2}}(-\frac{d^{2}}{dx^{2}}+\Delta^{4} x^{2}-2[ \frac{ \mathcal{H}_{m}''(\Delta x)}{\mathcal{H}_{m}(\Delta x)}-( \frac{ \mathcal{H}_{m}'(\Delta x)}{\mathcal{H}_{m}(\Delta x)})^{2}+\Delta^{2}])+\frac{\omega}{2} -E.
\end{equation}
Expressing $\tilde{h}^{(+)}$ and $\tilde{h}^{(-)}$ in the SUSY-like representation
\begin{equation}
\tilde{h}^{(+)}=\tilde{A^{\dagger}}\tilde{A},\quad \tilde{h}^{(-)}=\tilde{A}\tilde{A^{\dagger}}
\end{equation}
and noting that
\begin{equation}
\frac{1}{J}\tilde{h}^{(+)}=\frac{1}{\sqrt{J}}\tilde{A^{\dagger}}\frac{1}{\sqrt{J}}\tilde{A},\quad \frac{1}{J}\tilde{h}^{(-)}=\frac{1}{\sqrt{J}}\tilde{A}\frac{1}{\sqrt{J}}\tilde{A^{\dagger}}
\end{equation}
it readily follows
\begin{equation}
h^{(+)}-\frac{\omega}{2} =A^{\dagger}A,\quad h^{(-)}-\frac{\omega}{2}=A A^{\dagger}
\end{equation}
where we have defined
\begin{equation}
A=\frac{1}{\sqrt{J}}\tilde{A},\quad A^{\dagger}=\frac{1}{\sqrt{J}}\tilde{A^{\dagger}}.
\end{equation}
The operators $A$ and $A^{\dagger}$ are expressible in terms of the x-variable as
\begin{equation}
A=\frac{1}{\sqrt{J}}(\frac{1}{\Delta}\frac{d}{dx}+\frac{1}{\Delta}W(\Delta x)) \label{A}
\end{equation}
\begin{equation}
A^{\dagger}=\frac{1}{\sqrt{J}}(-\frac{1}{\Delta}\frac{d}{dx}+\frac{1}{\Delta}W(\Delta x)). \label{Ad}
\end{equation}
\par
%
%
Next, the superpotential $\tilde{W}(z) \rightarrow W(x)$ transforms as
\begin{equation}
\tilde{W}(z)=-\frac{d\tilde{\phi}(z)}{\tilde{\phi}(z)} \rightarrow -\frac{1}{\Delta}\frac{ \frac{d}{dx}\phi(\Delta x)}{\phi(x)}=\frac{1}{\Delta}W(\Delta x)
\end{equation}
\[ W(\Delta x) \equiv -\frac{\frac{d}{dx}\phi(\Delta x)}{\phi(\Delta x)}. \]
Since the seed solution is
\begin{equation}
\tilde{\phi_{m}}(z)=\mathcal{H}(z)e^{\frac{1}{2}z^{2}}
\end{equation}
which in the x-coordinate reads
\begin{equation}
\phi_{m}(\Delta x)= \mathcal{H}_{m}(\Delta x)e^{\frac{1}{2}\Delta^{2}x^{2}}
\end{equation}
it follows that $W(\Delta x)$ has the form
\begin{equation}
W(\Delta x)=-( \frac{\mathcal{H}_{m}'}{\mathcal{H}_{m}}+\Delta^{2} x).
\end{equation}
Hence from (\ref{A}) and (\ref{Ad}) we are led to the representations
\begin{equation}
A=\frac{1}{J}(\frac{1}{\Delta}\frac{d}{dx}-\frac{1}{\Delta} \frac{\mathcal{H}_{m}'}{\mathcal{H}}-\Delta x)
\end{equation}
\begin{equation}
A^{\dagger}=\frac{1}{J}(\frac{1}{-\Delta}\frac{d}{dx}-\frac{1}{\Delta} \frac{\mathcal{H}_{m}'}{\mathcal{H}}-\Delta x).
\end{equation}
The chain of Hamiltonians related via similarity transformations and the role of the ladder operators are summarized in Figure 1.
\begin{figure}
\begin{equation*}
\xymatrixcolsep{10pc}
\xymatrix{
( H^{(+)}, a^{\dagger},a)    \ar@{<->}[ddddd]^{\theta,\theta^{\dagger}}      \ar@{<->}[r]^{\rho,\rho^{-1}}            &        ( h^{(+)}, L,L^{\dagger})     \ar@{<->}[ddddd]^{A,A^{\dagger}}   \\
\\
\\
\\
\\
( H^{(-)})   \ar@{<->}[r]^{\rho,\rho^{-1}}            &       (  h^{(-)}, K,K^{\dagger})  }
\end{equation*}
\caption{Schematic presentation of the chain of Hamiltonians related via similarity transformations.}
\end{figure}
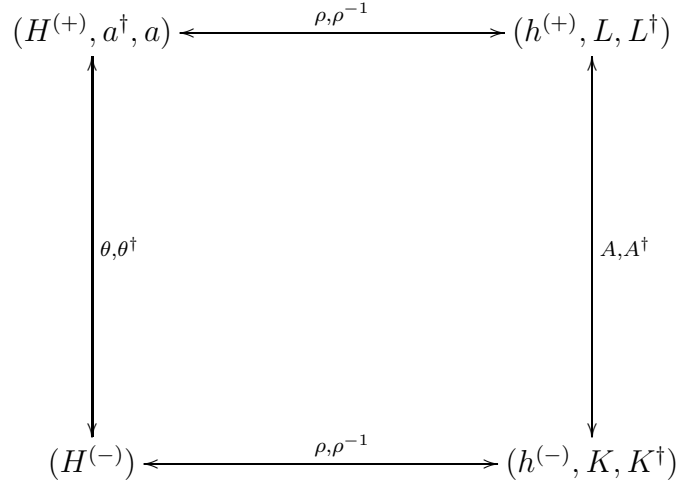
\par
%
It is interesting to point out that the Swanson oscillator and its generalized counterpart can be realized by the following pseudo supercharge operators:

\begin{equation}
\theta=\rho^{-1}A\rho=\frac{\lambda x}{\sqrt{J}\Delta}+\frac{1}{\sqrt{J}}\frac{1}{\Delta}\frac{d}{dx}+\frac{1}{\sqrt{J}}W(\Delta x)
\end{equation}

\begin{equation}
\theta^{\dagger}=\rho^{-1}A^{\dagger}\rho=-\frac{\lambda x}{\sqrt{J}\Delta}-\frac{1}{\sqrt{J}}\frac{1}{\Delta}\frac{d}{dx}+\frac{1}{\sqrt{J}}W(\Delta x)
\end{equation}
where
\begin{equation}
H^{(+)}-\frac{\omega}{2}=\theta^{\dagger}\theta,\quad H^{(-)}-\frac{\omega}{2}=\theta \theta^{\dagger}. \label{HpHm}
\end{equation}
In the last equation (\ref{HpHm}), $H^{(+)}$ represents the Swanson Hamiltonian. The Hamiltonians $H^{(+)}$ and $H^{(-)}$ can be represented in x-coordinate as
\begin{equation}
H^{(+)}=\frac{1}{J \Delta^{2}}(-\frac{d^{2}}{dx^{2}}-\lambda^{2}x^{2}-\lambda + \Delta^{2}+ \Delta^{4} x^{2}-2\lambda x \frac{d}{dx}+\Delta^{2} 2m)+\frac{\omega}{2}
\end{equation}
\begin{equation}
= -\frac{1}{2}(\omega -\alpha-\beta)\frac{d^{2}}{dx^{2}} + (\alpha-\beta)x\frac{d}{dx}+\frac{1}{2}(\alpha +\beta +\omega)x^{2}
\end{equation}
\[ (2m -1)\sqrt{\omega^{2}-4\alpha \beta}+\frac{(\alpha-\beta)}{2}+\frac{\omega}{2} \]
\begin{equation}
H^{(-)}=\frac{1}{J \Delta^{2}}(-\frac{d^{2}}{dx^{2}}-\lambda^{2}x^{2}-\lambda - \Delta^{2}+ \Delta^{4} x^{2}-2\lambda x \frac{d}{dx} \label{Hm}
\end{equation}
\[+ 2 (\frac{\mathcal{H}_{m}'}{\mathcal{H}_{m}})^{2} - 2 \frac{ \mathcal{H}_{m}''}{\mathcal{H}_{m}} +2 m \Delta^{2}    ) + \frac{\omega}{2} \]
\[= -\frac{1}{2}(\omega -\alpha-\beta)\frac{d^{2}}{dx^{2}} + (\alpha-\beta)x\frac{d}{dx}+\frac{1}{2}(\alpha +\beta +\omega)x^{2} \]
\[+ \frac{(\alpha-\beta)}{2}+\frac{\omega}{2} +\frac{1}{2}(\beta-\alpha) (-2 \frac{ \mathcal{H}_{m}''}{\mathcal{H}_{m}} + 2 (\frac{\mathcal{H}_{m}'}{\mathcal{H}_{m}})^{2})+(m+1)\frac{1}{2}\sqrt{\omega^{2}-4\alpha \beta}.\]
The above Hamiltonian speaks of the 1-step extension of the Swanson Hamiltonian $H^{(-)}$ as it is in the differential operators form and is connected to $h^{(-)}$ via a similarity transformation. Of the two Hamiltonians, $H^{(+)}$ coincide up to an additive constant with the non-Hermitian form obtained by inserting the coordinates representations of $a$ and $a^{\dagger}$ in the form (\ref{HS}) of $H^{s}$. (\ref{Hm}) is of course the supersymmetric partner to $H^{s}$. This result is new and of interest. For one thing, our derivation that the generalized counterpart of the Swanson Hamiltonian, as furnished by (\ref{Hm}), has an Hermitian equivalent Hamiltonian that is known to have wavefunctions involving exceptional orthogonal polynomials of type III-Hermite \cite{MarQue} is, to the best of our knowledge, a new result. For another, in addition to the various supercharges and ladder operators that appear in the $H^{s}$ , the Hermitian equivalent Hamiltonian also possesses ladder operators and related polynomial Heisenberg algebra. \par
%
%
The Hamiltonian $h^{(+)}$ has ladder operators
\begin{equation}
L=\frac{1}{\Delta}\frac{d}{dx}+\Delta x,\quad L^{\dagger}=-\frac{1}{\Delta}\frac{d}{dx}+\Delta x
\end{equation}
which satisfy
\begin{equation}
[h^{(+)},L]=-\frac{2}{J}L
\end{equation}
\begin{equation}
[h^{(+)},L^{\dagger}]=\frac{2}{J}L^{\dagger}
\end{equation}
\begin{equation}
L^{\dagger}L=J h^{(+)}-\frac{J \omega}{2}+JE-1
\end{equation}
\begin{equation}
LL^{\dagger}=J h^{(+)}-\frac{J\omega}{2}+JE+1.
\end{equation}
On the other hand the Hamiltonian $h^{(-)}$ is guided by the ladder operators
\begin{equation}
K=ALA^{\dagger},\quad K^{\dagger}=AL^{\dagger}A^{\dagger} \label{eqK}
\end{equation}
which satisfy
\begin{equation}
[h^{(-)},K]=-\frac{2}{J}K
\end{equation}
\begin{equation}
[h^{(-)},K^{\dagger}]=-\frac{2}{J}K^{\dagger}
\end{equation}
\begin{equation}
K^{\dagger}K=(J h^{(-)}-J\frac{1}{2}\omega+JE-1)(J h^{(-)}-\frac{J\omega}{2})(J h^{(-)}-\frac{J\omega}{2}-2)
\end{equation}
\begin{equation}
KK^{\dagger}=(J h^{(-)}-J\frac{1}{2}\omega+JE+1)(J h^{(-)}-\frac{J\omega}{2})(J h^{(-)}-\frac{J\omega}{2}+2).
\end{equation}
In Figure 1, $(h_{-},K,K^{\dagger})$ is represented as a result of mapping from $(h_{+},L,L^{\dagger})$ according to (\ref{eqK}).
\section{Two dimensional superintegrable generalisation}

Let us propose that a two-dimensional Swanson Hamiltonian will be superintegrable if its Hermitian equivalent form $h$ obtained from an overall similarity transformation is superintegrable i.e. if it allows two integrals of motion that are not only well defined but a polynomial of the momenta and be algebraically independent in addition to the Hamiltonian itself. In this regard we express the governing Hamiltonian as
\begin{equation}
\mathcal{H}=H_{1}+H_{2}=\omega_{1} a^{\dagger} a + \alpha_{1} a^{2} + \beta_{1} a^{\dagger 2} +\frac{1}{2}\omega_{1}
\end{equation}
\[+ \omega_{2} b^{\dagger} b + \alpha_{2} b^{2} + \beta_{2} b^{\dagger 2} +\frac{1}{2}\omega_{2} \]
in terms of two distinct sets of annihilation and creation operators namely $(a,a^{\dagger})$ and $(b,b^{\dagger})$. The latter are subject to the usual quantum conditions
\begin{equation}
[a,a^{\dagger}]=1,\quad [b,b^{\dagger}]=1.
\end{equation}
Then using $h=\rho_{1}\rho_{2}\mathcal{H}\rho_{1}^{-1}\rho_{2}^{-1}$ we have in the respective coordinates representations $x_{1}$ and $x_{2}$ the following result of superposition
\begin{equation}
h=h_{1}+h_{2}
\end{equation}
\[= -\frac{1}{2}(\omega_{1}-\alpha_{1}-\beta_{1})\frac{d^{2}}{dx_{1}^{2}}+\frac{1}{2}x_{1}^2(\frac{\omega_{1}^{2}-4 \alpha_{1}\beta_{1}}{\omega_{1}-\alpha_{1}-\beta_{1}})\]
\[-\frac{1}{2}(\omega_{2}-\alpha_{2}-\beta_{2})\frac{d^{2}}{dx_{2}^{2}}+\frac{1}{2}x_{2}^2(\frac{\omega_{2}^{2}-4 \alpha_{2}\beta_{2}}{\omega_{2}-\alpha_{2}-\beta_{2}}).\]

Ladder operators of the types described previously exist in both the coordinates and are provided by the representations such as

\begin{equation}
A=h_{1}-h_{2}
\end{equation}
and the pair

\begin{equation}
L_{1}=\frac{1}{\Delta_{1}}\frac{d}{dx_{1}}+\Delta_{1}x_{1}
\end{equation}
\begin{equation}
L_{2}=\frac{1}{\Delta_{2}}\frac{d}{dx_{2}}+\Delta_{2}x_{2}.
\end{equation}
Thus the following construction proposed recently \cite{Mar2}
\begin{equation}
I_{-}= (L_{1})^{n_{1}}(L_{2}^{\dagger})^{n_{2}}
\end{equation}
\begin{equation}
I_{+}= (L_{1}^{\dagger})^{n_{1}}(L_{2})^{n_{2}}
\end{equation}
holds subject to the constraint
\begin{equation}
n_{1} \sqrt{\omega_{1}-4\alpha_{1}\beta_{1}}=n_{2} \sqrt{\omega_{2}-4\alpha_{2}\beta_{2}} \label{constraints}
\end{equation}
where $n_{1}$ and $n_{2}$ are integers. A family of superintegrable systems is thus provided by
\begin{equation}
B_{1}=L_{-}-L_{+},\quad B_{2}=L_{-}+L_{+}.
\end{equation}
Noting that the integral $A$ with any of the type $I_{-}$, $L_{+}$, $B_{1}$ or $B_{2}$ will generate three integrals of motion algebraically independent of the Hamiltonian $H$, we observe that the same can be done for the generalized Swanson Hamiltonian from its 1 step extension.
Expressing $\mathcal{H}$ in terms of the coordinates $x_{1}$ and $x_{2}$
\begin{equation}
\mathcal{H}=H_{1}+H_{2}= -\frac{1}{2}(\omega_{1} -\alpha_{1}-\beta_{1})\frac{d^{2}}{dx_{1}^{2}} + (\alpha_{1}-\beta_{1})x_{1}\frac{d}{dx_{1}}+\frac{1}{2}(\alpha_{1} +\beta_{1} +\omega_{1})x_{1}^{2}
\end{equation}
\[+ \frac{(\alpha_{1}-\beta_{1})}{2}+\frac{\omega_{1}}{2} +\frac{1}{2}(\beta_{1}-\alpha_{1}) (-2 \frac{ \mathcal{H}_{m_{1}}''}{\mathcal{H}_{m_{1}}} + 2 (\frac{\mathcal{H}_{m_{1}}'}{\mathcal{H}_{m_{1}}})^{2})+(m_{1}+1)\frac{1}{2}\sqrt{\omega_{1}^{2}-4\alpha_{1} \beta_{1}}\]
\[-\frac{1}{2}(\omega_{2} -\alpha_{2}-\beta_{2})\frac{d^{2}}{dx_{2}^{2}} + (\alpha_{2}-\beta_{2})x_{2}\frac{d}{dx_{2}}+\frac{1}{2}(\alpha_{2} +\beta_{2} +\omega_{2})x_{2}^{2} \]
\[+ \frac{(\alpha_{2}-\beta_{2})}{2}+\frac{\omega_{2}}{2} +\frac{1}{2}(\beta_{2}-\alpha_{2}) (-2 \frac{ \mathcal{H}_{m_{2}}''}{\mathcal{H}_{m_{2}}} + 2 (\frac{\mathcal{H}_{m_{2}}'}{\mathcal{H}_{m_{2}}})^{2})+(m_{2}+1)\frac{1}{2}\sqrt{\omega_{2}^{2}-4\alpha_{2} \beta_{2}}\]
and using $h=\rho_{1}\rho_{2}H\rho_{1}^{-1}\rho_{2}^{-1}$ we find that
\begin{equation}
h=h_{1}+h_{2} \label{2Dh}
\end{equation}
\[ = \frac{1}{J_{1} \Delta_{1}^{2}}(-\frac{d^{2}}{dx_{1}^{2}}+\Delta_{1}^{4} x_{1}^{2}-2[ \frac{ \mathcal{H}_{m_{1}}''(\Delta x_{1})}{\mathcal{H}_{m_{1}}(\Delta_{1} x_{1})}-( \frac{ \mathcal{H}_{m_{1}}'(\Delta_{1} x_{1})}{\mathcal{H}_{m_{1}}(\Delta_{1} x_{1})})^{2}+\Delta_{1}^{2}])+\frac{\omega_{1}}{2} -E_{1} \]
\[ + \frac{1}{J_{2} \Delta_{2}^{2}}(-\frac{d^{2}}{dx_{2}^{2}}+\Delta_{2}^{4} x_{2}^{2}-2[ \frac{ \mathcal{H}_{m_{2}}''(\Delta_{2} x_{2})}{\mathcal{H}_{m_{2}}(\Delta_{2} x_{2})}-( \frac{ \mathcal{H}_{m_{2}}'(\Delta_{2} x_{2})}{\mathcal{H}_{m_{2}}(\Delta_{2} x_{2})})^{2}+\Delta_{2}^{2}])+\frac{\omega_{2}}{2} -E_{2} .\]
Eq(\ref{2Dh}) gives the full algebraic structure of $h$ which is not merely a multiplication of two commuting one-dimensional systems as is evidenced by the connection of a underlying constraint (\ref{constraints}). Such a Hermitian equivalence of a superintegrable system has not been explored before. The overall similarity transformation applies to the integrals of the Hermitian equivalent Hamiltonian also allows to obtain operators commuting with the two-dimensional non Hermitian operator. \par
%
%

 %
 %
 \section{CONCLUSION}

 In this article we have discussed the possibility of the Swanson Hamiltonian $H^{s}$ admitting a one step rational extension and determined the same by using the standard techniques of supersymmetric quantum mechanics. To this end we have exploited the various representations of the ladder operators using a seed solution of the harmonic oscillator which is given in terms of the pseudo-Hermitian polynomials as the basic starting point. We have in this way formulated a flow-chart of such operators that has guided us to move from one system of a SUSY-like pair to another that are related by a similarity transformation and finally closing on $H^{s}$ itself. We have also proposed a two-dimensional generalization of the Swanson Hamiltonian and constructed appropriate integrals of motion of the Hermitian equivalent to establish superintegrabilty of such a system. Recently, some two-dimensional harmonic oscillator and pseusoboson have been constructed \cite{Bar}.

 \setcounter{equation}{0}

\par
%
%
\section*{ACKNOWLEDGMENTS}

The research of I.\ M.\ was supported by the Australian Research Council through Discovery Early Career Researcher Award DE130101067. B.B. acknowledges the warm hospitality of Dr Ian Marquette during his stay at the University of Queensland. We also thank the anonymous referees for making a number of constructive comments. \par
%
%

%
%

\end{document}